\documentclass[pra,aps,showpacs,twocolumn,floatfix]{revtex4}
\usepackage{graphicx}
\usepackage[ansinew]{inputenc}
\usepackage{array}
\usepackage{color}
\usepackage{amsmath}
\usepackage{amsxtra}
\usepackage{amstext}
\usepackage{amssymb}
\usepackage{latexsym}
\usepackage{dsfont}
\begin{document}

\title{Trapping and cooling a mirror to its quantum mechanical ground state}

\author{M. Bhattacharya and P. Meystre}
\affiliation{Department of Physics, The University of Arizona,
Tucson, Arizona 85721}

\date{\today}

\begin{abstract}
We propose a technique aimed at cooling a harmonically oscillating
mirror to its quantum mechanical ground state starting from room
temperature. Our method, which involves the two-sided irradiation
of the vibrating mirror inside an optical cavity, combines several
advantages over the two-mirror arrangements being used currently.
For comparable parameters the three-mirror configuration provides
a stiffer trap for the oscillating mirror. Furthermore it prevents
bistability from limiting the use of higher laser powers for
mirror trapping, and also partially does so for mirror cooling.
Lastly, it improves the isolation of the mirror from classical
noise so that its dynamics are perturbed mostly by the vacuum
fluctuations of the optical fields. These improvements are
expected to bring the task of achieving ground state occupation
for the mirror closer to completion.
\end{abstract}


\maketitle

The observation of quantum dynamics in truly macroscopic objects
is a challenging task that, although not yet achieved, has begun
to look more and more feasible
\cite{gigan2006,kleckner2006,arcizet2006,schliesser2006,solstate}
as a result of recent experimental advances that include novel
cooling techniques, progress in nanofabrication, and the improved
control of decoherence. This is an exciting prospect, as it would
enable us to explore the quantum-classical boundary
\cite{leggett2002} as well as to test quantum mechanics in an
entirely new regime \cite{marshall2003}. The implementation of
characteristically quantum mechanical phenomena such as
entanglement and superposition at a macroscopic scale
\cite{mancini2002} has broad implications, for instance in the
impending merger of quantum optics with condensed matter physics
\cite{hansch2007}. It also signals direct technological benefits
for areas like quantum measurement \cite{caves1980} and
communication \cite{vitali2007}. Additional potential applications
include the interferometric detection of gravitational waves
\cite{courty2003} and atomic force microscopy \cite{walls1994}.

\begin{figure}
\includegraphics[width=0.48 \textwidth]{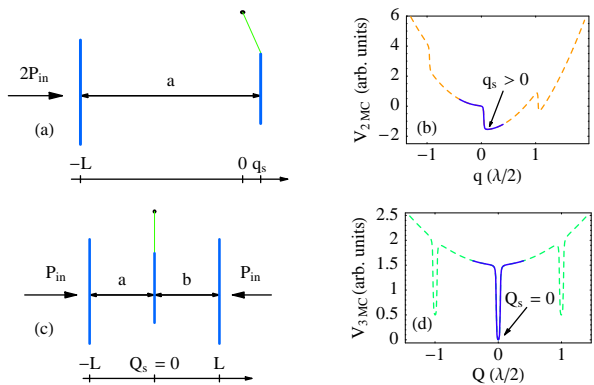}
\caption{\label{fig:cavitypic}(Color online).(a) Two-mirror cavity (2MC), with
$q_{s}$ the static recoil due to radiation pressure. (b) The
corresponding mirror trapping potential in the static limit. The dashed
line shows the full classical solution, while the solid line is the
$\omega=0$ limit of the single-mode quantum theory. (c) Three-mirror
cavity (3MC). (d) The corresponding mirror trapping potential in the static
limit. The dashed line shows the full classical solution, while the
solid line is the $\omega=0$ limit of the two-mode quantum theory. The 
parameters labelling the figures are defined in the text.}
\end{figure}

A promising route to these objectives seems to be through the use
of optomechanical systems, particularly of mirror cavities coupled
to laser radiation \cite{dorsel1983,braginsky2002,cohadon1999}.
Experimentally two-mirror cavities have been used, with one mirror
held fixed and another mounted on a spring of frequency
$\Omega_{M}$ and damped at a rate $\Gamma_{M}$
[Fig.\ref{fig:cavitypic}(a)]. It has been demonstrated that
radiation pressure in the cavity can change these quantities to
new values $\Omega_{\rm eff}$ and $\Gamma_{\rm eff}$
\cite{karrai2004}. To bring the mirror to its quantum mechanical
ground state starting from temperature $T$ the number of quanta
\cite{karrai2004}
\begin{equation}
 \label{eq:quanta}
n= \frac{k_{B}T}{\hbar \Omega_{M}} \frac{\Gamma_{M}}{\Gamma_{\rm
eff}} \left(\frac{\Omega_{M}}{\Omega_{\rm eff}}\right)^{3},
\end{equation}
have to be reduced to $ < 1$. Here $k_{B}$ and $\hbar$ are
Boltzmann's and Planck's constants.

Most progress so far has been achieved by using the technique of
\textit{cold damping} where the thermal Brownian motion heating up
the oscillator is countered by the effect of laser radiation,
which increases the mirror damping while introducing very little noise
\cite{pinard2000}. As a result the ratio $\Gamma_{M}/\Gamma_{\rm
eff}$ can be decreased to about $10^{-4}$ either passively or by
using active feedback cooling. This yields $n \sim 10^{4}$ for
typical parameters even if the base temperature is cryogenically
reduced to $T \sim 4$K \cite{poggio2007}. We mention here that $n
\sim 25$ has also been attained by coupling nanomechanical
resonators to single-electron transistors in the solid state
\cite{solstate}. Cold damping works best for oscillating mirrors
with high mechanical frequencies and quality factors and
high-finesse cavities. However mirror heating effects due to
absorption \cite{gigan2006} and the onset of bistability
\cite{dorsel1983} impose a limit to the laser powers that can be
used. Improving the cavity finesse may reduce mirror heating, but
it also lowers the power threshold for bistability. We also note
that cold damping works for laser fields tuned \textit{below} the
cavity resonance \cite{gigan2006}. For the conditions just
described there is only a small downshift of the mechanical
resonance frequency, i.e. $\Omega_{M}/\Omega_{eff} \sim 1$.

On the other hand, Eq.(\ref{eq:quanta}) indicates that if the
spring could be made stiffer
\cite{pm1985,khalili2001,vogel2003,sheard2004} even by an order of
magnitude, the number of quanta could be lowered by three orders
of magnitude. However this requires the radiation to be tuned
\textit{above} the cavity resonance, which precludes cold damping.
The resolution to this problem has been demonstrated very recently
\cite{corbitt2007} in an experiment that demonstrated the almost
independent control of $\Gamma_{\rm eff}$ and $\Omega_{\rm eff}$
with the use of two radiation fields of different powers and
detunings. One laser field was used to stiffen the spring while
the other achieved cold damping, leading to $n \sim 10^{3}$
starting from room temperature. This method also makes it possible
to work with weak mechanical attachment of mirrors (small
$\Omega_{M}$), implying a lower coupling to thermal noise and
longer thermal decoherence times. Forming tighter traps therefore
seems to be a way to surmount the current limitations of cold
damping in reaching the mirror ground state.

Accounting for these observations, the present Letter proposes and
analyzes a mirror cooling and trapping configuration that provides
stiffer traps, largely removes the problem of bistability and
better isolates the mirror from thermal noise. These improvements
relax considerably the difficulties in cooling the macroscopic
mirror to its quantum mechanical ground state. The arrangement
that we propose involves placing a mirror with two perfectly
reflecting surfaces inside a cavity where the other mirrors are
transmissive but not movable [Fig.\ref{fig:cavitypic}(c)]. The
cavity is then driven by two different laser fields, one for
cooling and the other for trapping from both sides.
The three-mirror cavity (3MC) configuration leads to potential
wells for the movable mirror that are tighter and more harmonic
than is the case for the two-mirror cavity (2MC) as illustrated in
Fig.\ref{fig:cavitypic}. Such a geometry was proposed many years
ago and shown to provide superior mirror confinement, in addition to
being more stable than the 2MC when subjected to Brownian noise
\cite{pm1985}. However the treatment underlying that proposal was
classical, constructed using solutions to the classical Maxwell's
equations for the optical fields in the Fabry-P\'{e}rot, the trap
frequencies were derived in the static limit, and the noise
analysis did not account for fluctuations in the laser fields.

Here we show that tight confinement of the movable mirror in the
3MC is retained in a dynamical quantum mechanical picture that
fully accounts for damping and noise. As expected the classical
results previously obtained are regained from the quantum
treatment in the appropriate limit [see Figs.\ref{fig:cavitypic}(b)
and (d)]. The key point is then that the higher symmetry of the
3MC allows us to significantly improve the trap stiffness and to
partially remove the bistability that is a dominant feature of the
2MC configuration. These conclusions follow from the fact that in
the 3MC the movable mirror does not possess the static recoil due
to radiation pressure that follows from the lack of symmetry in
the 2MC.

We now proceed to outline a derivation of our results for the 3MC,
and contrast them to a similar treatment of the 2MC for
comparison. For our purposes it is adequate to consider external
pumping fields at a single frequency, although in practice two
frequencies have to be used as discussed above. We first consider
cavities without control and subsequently include electronic
feedback in our discussion.

The Hamiltonian for the 3MC shown in Fig.\ref{fig:cavitypic}(c)
can be written as  \cite{mancini2002}
\begin{equation}
\label{eq:Ham}
H = \hbar \omega_{c}(a^{\dagger}a+b^{\dagger}b)+\frac{P^{2}}{2M}
+\frac{1}{2}M \Omega_{m}^{2}Q^{2}
-\hbar\xi(a^{\dagger}a-b^{\dagger}b)Q,
\end{equation}
where $a$ and $b$ ($a^\dagger$ and $b^\dagger$) are the bosonic
annihilation (creation) operators for the electromagnetic modes of
equal frequency $\omega_{c}$ in the left and right sub-cavities of
length $L$, $\xi=\omega_{c}/L$ is the optomechanical coupling
constant and $P, Q$ are the momentum and displacement operators of the
middle mirror of mass $M$. It is assumed in the following that $Q
\ll \lambda$, the optical wavelength.

The first term in the Hamiltonian corresponds to the energy of the
optical fields in the two sub-cavities, the next two terms account
for the energy of the oscillating middle mirror and the last term
describes the effect of radiation pressure on that mirror. We
include damping and noise in our treatment via the familiar
input-output theory. The Hamiltonian (\ref{eq:Ham}) yields then the
quantum Langevin equations \cite{gardinerbook}
\begin{eqnarray}
\label{eq:QLE1}
 \begin{array}{ll}
\dot{a}= &-\left[i(\Delta-\xi Q)+\frac{\gamma}{2}\right]a+\sqrt{\gamma} a^{\rm in},\\
\dot{b}= &-\left[i(\Delta+\xi Q)+\frac{\gamma}{2}\right]b+\sqrt{\gamma} b^{\rm in},\\
\dot{Q}= & P/M,\\
\dot{P}= & -M\Omega_{M}^{2}Q+\hbar \xi (a^{\dagger}a-b^{\dagger} b)
-\frac{\Gamma_{M}}{2M}P+\epsilon^{\rm in},\\
\end{array}
\end{eqnarray}
where $\gamma$ is the decay rate of both the left and right
sub-cavities, and $\Delta=\omega_{c}-\omega_{L}$ is the detuning introduced
to account for the difference between 
the cavity resonance and the laser frequency $\omega_{L}$. The operators $a^{\rm in}$ and
$b^{\rm in}$ are noise operators describing the fields incident on
the cavity. Their mean values $\langle a^{\rm in}(t)\rangle
=a_{s}^{\rm in}, \langle b^{\rm in}(t) \rangle =b_{s}^{\rm in}$
describe classical fields pumping the cavities, and their
fluctuations are assumed to be delta-correlated
$\langle \delta a^{\rm in}(t) \delta a^{\rm in,\dagger}(t')
\rangle=
\langle \delta b^{\rm in}(t) \delta b^{\rm in,\dagger}(t') \rangle=\delta (t-t'),$
and add vacuum noise to the cavity modes. Finally, the operator
$\epsilon^{\rm in}$ describes the Brownian noise resulting from
the coupling of the middle mirror to the environment. Its mean
value is zero, and its fluctuations at temperature $T$ are characterized
by the correlation 
$\langle \delta \epsilon^{\rm in}(t) \delta \epsilon^{\rm in}(t') \rangle=
\frac{\Gamma_{M}}{2M}\int_{-\infty}^{\infty} \frac{d\omega}{2\pi}e^{-i\omega(t-t')}\hbar \omega
\left[1+ \coth \left(\frac{\hbar \omega}{2k_{B}T}\right)\right]$ \cite{gardinerbook}.

The steady-state values of the dynamical variables can be found by setting the
time-derivatives in Eq.(\ref{eq:QLE1}) to zero. This yields
\begin{equation}
\label{eq:sstate} a_{s}=b_{s}\equiv
f_{s}=\frac{\sqrt{\gamma}f_{s}^{\rm
in}}{\left(\frac{\gamma^{2}}{4}+\Delta^{2}\right)^{1/2}},
\,\,\,\,Q_{s}=0, \,\,\,\,P_{s}=0,
\end{equation}
where $f_{s}^{\rm in}=|a_{s}^{\rm in}|=|b_{s}^{\rm in}|$ is the
amplitude of both pumping fields. The phases of these fields have
been independently chosen such that $f_{s}$ is real. This can be
done without loss of generality as our model is insensitive to the
mutual (in)coherence of the two pumping fields. For later use we
define the input power from either side as $P_{\rm in}=\hbar
\omega_{c}|f_{s}^{\rm in}|^{2}$.

It can be shown from the steady-state equations that for $\Delta
<0$, which corresponds to trapping and heating of the middle
mirror, that $Q_{s}=0$ is always the only real solution and hence
\textit{there is no multistability in the position of the mirror}.
This is in contrast to the 2MC where bistability is expected for
$|\Delta| \geqslant \sqrt{3}\gamma/2$ \cite{mancini1994}, and has
been observed \cite{pm1985}. Physically, this is a result of the
fact that the cavity length changes with $P_{in}$ in a fashion
mathematically similar to the effect of a Kerr medium in
conventional bistability. The implication for mirror control is
that higher laser powers can be used in the 3MC than in the 2MC to
strengthen the trapping effects of radiation pressure, and that
these powers are now limited only by absorption heating effects.

The stiffening due to the trapping laser field also improves the
situation with regard to the bistability that occurs due to the
cooling (and anti-trapping) laser field at $\Delta >0$: It takes
more power to make a stiff mirror bistable, hence the trapping
field raises the power threshold for the onset of bistability due
to the cooling field. We show below that the 3MC can provide
stiffer traps than the 2MC. This implies that higher laser power
can be used for cooling in the 3MC than in the 2MC.

We mention here that depending on the laser powers used for
cooling and trapping fields, their respective detunings have to be
chosen to satisfy requirements of dynamic stability and are not
completely independent \cite{corbitt2007}. Typically, for the
relatively high powers that we consider here stability is
achievable for large detunings for the trapping laser field and
small detunings of the cooling field. It is therefore not always
possible to stay within the window of detunings $|\Delta|
\leqslant \sqrt{3}\gamma/2$ where the 2MC dynamics are
monostable, and bistability inevitably becomes an issue.

To investigate the effect of noise on the steady-state values
Eq.~(\ref{eq:sstate}) we consider each operator in
Eq.~(\ref{eq:QLE1}) as the sum of its corresponding $c-$number
mean value from Eq.~(\ref{eq:sstate}) and a small fluctuation
around that value. For example, $a=a_{s}+\delta a$. Using these
definitions in Eq.~(\ref{eq:QLE1}), eliminating the steady-state
conditions and linearizing the remaining equations in the
fluctuations, we can write compactly 
\begin{equation}
\label{eq:linear}
\dot{\bar{u}}(t)= A\bar{u}(t)+\bar{n}(t).
\end{equation}
Here the input noise vector is given by
$\bar{n}(t)=(\sqrt{\gamma}X_{a}^{\rm in},\sqrt{\gamma}Y_{a}^{\rm
in},\sqrt{\gamma}X_{b}^{\rm in},\sqrt{\gamma}Y_{b}^{\rm
in},0,\delta \epsilon^{\rm in})$ and the vector of fluctuations in
the dynamical variables by $\bar{u}(t)=(\delta X_{a},\delta
Y_{a},\delta X_{b},\delta Y_{b},\delta Q,\delta P)$, where we have
redefined the field fluctuations as $\delta X_{a} =(\delta
a+\delta a^{\dagger})/\sqrt{2}$ and $\delta Y_{a} =(\delta
a-\delta a^{\dagger})/i\sqrt{2}$, for example. Further,
\begin{equation}
\label{eq:RHmatrix} A =
\begin{pmatrix}
-\frac{\gamma}{2}  &  \Delta             &  0                    &       0            &   0                   &   0\\
 -\Delta            & -\frac{\gamma}{2}   &  0                    &        0           & \sqrt{2}\hbar\xi f_{s}     &   0\\
      0             & 0                   & -\frac{\gamma}{2}    & \Delta             &   0                   &          0\\
 0                  &    0                & -\Delta               & -\frac{\gamma}{2}  & -\sqrt{2}\hbar\xi f_{s}   &  0\\
0               &    0             &     0                   &   0                   & 0                     &  \frac{1}{M}\\
\sqrt{2}\hbar\xi f_{s} &    0       &-\sqrt{2}\hbar\xi f_{s} &   0 & -M \Omega_{M}^{2} &-\frac{\Gamma_{M}}{2M}\\
\end{pmatrix}.
\end{equation}
The solutions to Eq.(\ref{eq:linear}) are stable when none of the
eigenvalues of the matrix $A$ has a positive real part. This
condition can be formalized using the Routh-Hurwitz criterion
\cite{dejesus1987}; the corresponding inequalities are too
involved to be reproduced here, but we have verified that they
hold for the parameter values considered in this Letter.

In order to solve Eq.(\ref{eq:linear}) we Fourier transform it
into a set of algebraic equations. We then find the mechanical
susceptibility $\chi (\omega)$ of the middle mirror by solving for
$\delta Q(\omega) =  \chi (\omega) F_{T}(\omega)$
\cite{cohadon1999}, where $F_{T}$ is the total force on the middle
mirror and includes the Brownian and radiation forces. The
susceptibility has the form of a Lorentzian $\chi^{-1} (\omega) =
M(\Omega_{\rm eff}^{2}-\omega^{2})-i\Gamma_{\rm eff}\omega$
\cite{cohadon1999}, where the effective oscillation frequency and
damping for the mirror are given by
\begin{eqnarray}
\label{eq:eff}
\begin{array}{ll}
\Omega_{\rm eff}^{2}=&\Omega_{M}^{2}-\frac{4\xi \gamma P_{\rm
in}}{ML} \frac{\Delta}{\Delta^{2}+\frac{\gamma^{2}}{4}}
\left[\frac{\Delta^{2}-\omega^{2}+\frac{\gamma^{2}}{4}}
{(\Delta^{2}-\omega^{2}+\frac{\gamma^{2}}{4})^{2}+\gamma^{2}\omega^{2}}\right],\\
\Gamma_{\rm eff}=&\frac{\Gamma_{M}}{2}+\frac{4\xi \gamma P_{\rm
in}}{ML} \frac{\Delta}{\Delta^{2}+\frac{\gamma^{2}}{4}}
\left[\frac{\gamma}
{(\Delta^{2}-\omega^{2}+\frac{\gamma^{2}}{4})^{2}+\gamma^{2}\omega^{2}}\right].\\
\end{array}
\end{eqnarray}
In the static limit $(\omega=0)$ $\Omega_{\rm eff}$ reduces to the
classical expression for the frequency of the mirror motion in the
lowest minima of the potential shown in
Fig.\ref{fig:cavitypic}(d).

In order to carry out a comparison we calculate the corresponding
effective quantities $\omega_{\rm eff}, \gamma_{\rm eff}$ for the
2MC in a similar way and with identical parameters, but pumping
with $2P_{\rm in}$ [Fig.\ref{fig:cavitypic}(a)]. (We preserve the 
notation for the fixed mirror
and the optical fields, but use lower case for the movable mirror
variables.) In this paper we restrict ourselves to comparing the
frequencies $\omega_{\rm eff}$ and $\Omega_{\rm eff}$ for brevity.

The expressions for $\omega_{\rm eff}$ and $\gamma_{\rm eff}$ turn
out to be identical to Eq.~(\ref{eq:eff}) except for the
replacement of $\Delta$ by the effective detuning
\begin{equation}
\label{eq:detuning2}
\Delta'=\Delta-\xi q_{s}.
\end{equation}
Here $\xi q_{s}$ is the radiation pressure induced detuning due to
the non-zero equilibrium position for the mirror, $q_{s}$
[Fig.\ref{fig:cavitypic}(a)]. This displacement is determined from
the steady-state balance of the radiation and spring forces on the
mirror
\begin{equation}
\label{eq:forcebalance} \frac{2\gamma P_{\rm
in}/L}{(\frac{\gamma}{2})^{2}+(\Delta-\xi
q_{s})^{2}}-m\omega_{m}^{2}q_{s}=0,
\end{equation}
which is a cubic in $q_{s}$. Below the onset of bistability
Eq.~(\ref{eq:forcebalance}) has a single real root that is always
positive and grows with input power $(q_{s} \sim P_{\rm
in}^{1/3})$. For high $P_{\rm in}$ the vibrational properties of
the mirror are dominated by the optical fields, that is, the
contributions due to $\Omega_{M}$ and $\Gamma_{M}$ in
Eq.(\ref{eq:eff}) are negligible compared to the remaining terms
and similarly for $\omega_{\rm eff},\gamma_{\rm eff}$ .

Since the spring is soft we examine the frequency ratio at
$(\omega \ll \gamma/2)$ \cite{corbitt2007}. This limit essentially
yields the classical potentials derived earlier \cite{pm1985} and
shown in Figs.\ref{fig:cavitypic}(b) and (d). The frequency ratio
    \begin{equation}
    \frac{\Omega_{\rm eff}}{\omega_{\rm eff}} \sim
    \left(\frac{\Delta}{\Delta'}\right)^{1/2}
    \left(\frac{\Delta'^{2}+\gamma^{2}/4}
    {\Delta^{2}+\gamma^{2}/4}\right)
    \end{equation}
implies a very significant improvement in trapping in the 3MC as
compared to the 2MC since $|\Delta| < |\Delta'|$ --- recall that
the number of quanta $n$ scales as $\Omega_{\rm eff}^{-3}$, see
Eq.~(\ref{eq:quanta}). For $M=$1mg, $\Omega_{M}=2\pi \times
100$Hz, $L=2.5$cm, $\gamma=10$MHz, $\lambda = 1064$nm, $P_{\rm
in}=1$mW and $\Delta =-\gamma/2$, $\Delta' \sim -4.5\gamma$ and
$\left(\Omega_{\rm eff}/\omega_{\rm eff}\right) \sim 10$. The
actual frequencies are 1.6 $\Omega_{M}$ and 18$\Omega_{M}$ for the
2MC and 3MC configurations, respectively.

It is clear that accounting for $\left(\Omega_{\rm
eff}/\omega_{\rm eff}\right)$ in Eq.(\ref{eq:quanta}) and assuming
similar cold damping performance, we can achieve numbers of quanta
three orders of magnitude smaller in the 3MC than in the 2MC
configuration. Specifically, for our cavity parameters and the
length $L$ shortened to $L=1$cm, trapping light at $P_{\rm
in}^{t}=400$mW, $\Delta^{t}=-2.5\gamma$ and cooling light at
$P_{\rm in}^{c}=5$mW, $\Delta^{c}=\gamma/2$, it is possible to
achieve $n < 1$ when starting from room temperature, that is, the
mirror can be cooled to its ground state of motion.

Having demonstrated the advantages of the 3MC configuration, we
now discuss limitations and ramifications of our proposal. It is
particularly relevant to consider here the role of feedback as it
can influence the mirror stiffness. In the 2MC case, if the laser
is locked to the cavity \cite{arcizet2006}, or if a transducer is used to control the
`fixed' mirror and keep the cavity length constant \cite{gigan2006}, the movable
mirror feels the spring force from both sides, but the radiation pressure 
force only from one side yielding the asymmetric potential
of Fig.\ref{fig:cavitypic}(b).

However if the transducer (such as a piezoelectric crystal or
magnetic coil) is mounted on the movable mirror \cite{corbitt2007}, 
then feedback
governing the length of the cavity provides a force equal and
opposite to the radiation pressure force in the cavity (within the
bandwidth of the loop), yielding a potential similar to the 3MC
case shown in Fig.\ref{fig:cavitypic}(d). But such a force brings
with it classical noise, since it essentially corresponds to
(one-sided) stiffening of the mechanical connection of the mirror.
Moreover, with nanomechanical oscillators being used in many
experiments \cite{gigan2006,kleckner2006,arcizet2006,schliesser2006} 
the possibility of using purely optical control
seems more attractive for miniaturization as well as scaling. The
3MC configuration that we propose is in a sense a form of feedback
where the noise contribution is quantum mechanical, as classical
noise can be suppressed in the laser light. The same effect could
be achieved with an independent laser beam used from the right on
the 2MC, but it would require impracticably high powers. The
addition of an extra cavity allows for lower laser power to build
up to the same effect. With the higher optical stiffness that can
be achieved with the 3MC the role of classical noise is therefore
reduced further.

In conclusion we have proposed a laser cooling and trapping
geometry that provides stiffer traps, removes bistability
partially, and isolates the mirror further from classical
mechanical noise. We believe that it offers a promising
experimental route for confining and cooling a macroscopic mirror
near its quantum mechanical ground state than heretofore proposed
or achieved.

This work is supported in part by the US Office of Naval Research,
by the National Science Foundation, by the US Army Research
Office, by the Joint Services Optics Program, and by the National
Aeronautics and Space Administration. We would like to thank H. Uys
for stimulating discussions.

\end{document}